\documentclass[iicol]{sn-jnl}

\usepackage{epsfig,amssymb,amsfonts,amsmath,mathtools,bm,color,graphicx,orcidlink,bm,braket,multirow,dsfont,mathtools,xcolor,slashed}

\RequirePackage[numbers,sort&compress]{natbib}

\renewcommand{\vec}[1]{\mbox{\boldmath$#1$\unboldmath}}
\newcommand{\Tch}{T_{\rm ch}}
\newcommand{\mq}{m_q}

\newcommand{\be}{\begin{equation}}
\newcommand{\ee}{\end{equation}}
\newcommand{\bea}{\begin{eqnarray}}
\newcommand{\eea}{\end{eqnarray}}
\newcommand{\beas}{\begin{eqnarray*}}
\newcommand{\eeas}{\end{eqnarray*}}

\newcommand{\vep}{{\bm p}}

\newcommand{\veq}{{\bm q}}

\newcommand{\vex}{{\bm x}}
\newcommand{\vey}{{\bm y}}

\newcommand{\muIR}{\mu_{\scalebox{0.6}{{\rm IR}}}}

\def\vec#1{{\bm #1}}

\graphicspath{{Figs.dir/}}

\begin{document}

\title{On the  origin of the $N_c^1$ scaling in the confined
but chirally symmetric  phase at high T.}

\author[1]{L. Ya. Glozman}\email{E-mail: leonid.glozman@uni-graz.at}

\affil{\orgdiv{Institute of Physics}, \orgname{University of Graz}, \orgaddress{ \postcode{8010}, \city{Graz}, \country{Austria}}}


\date{}

\abstract{
There is lattice evidence that
the QCD  matter above the chiral restoration temperature $T_{ch}$ and below the
deconfinement temperature $T_d$, called stringy fluid, is characterized
by  approximate chiral spin symmetry, which is a symmetry of confinement
in QCD with light quarks. The energy density, pressure and entropy density in the stringy fluid scale as $N_c^1$, which is in contrast to the $N_c^0$ scaling
 in the hadron gas and to the $N_c^2$ scaling in the quark-gluon plasma.
 Here we clarify the  origin of the $N_c^1$ scaling.  We employ a solvable field-theoretical large $N_c$ chirally symmetric and confining 
 model. In vacuum the confining potential induces a spontaneous breaking
 of chiral symmetry. The mesons are spatially localized  states of quarks and antiquarks. Still
  in the confining regime 
  the system undergoes the chiral restoration phase transition at $T_{ch}$  because of  
  Paili blocking of the quark levels required for the existence of the quark
 condensate, by the thermal excitation of quarks and antiquarks. The same
 Paili blocking leads to a delocalization of the color singlet low-spin meson-like
 states that become infinitely large in the chiral limit.  
 Consequently the stringy fluid  represents a  very dense medium of the overlapping huge color-singlet low-spin quark-antiquark systems.
  The Bethe-Salpeter equation that determines the rest-frame excitation
 energies of  the color-singlet quark-antiquark system  is
 $N_c$-independent both in vacuum and in the medium in the confining regime. The excitation energy
  of the quark-antiquark color-singlet systems scales as $N_c^0$, i.e. as
  meson mass in vacuum.
  The $N_c^1$ scaling
  of the energy density in the stringy fluid is provided by the  fluctuations
  of the color-singlet quark-antiquark systems. 
}

\maketitle

\section{Introduction}

Experimentally,   the hot QCD matter
above the chiral restoration crossover around  $\Tch \simeq 155$ MeV differs radically from  both a dilute hadron gas at low temperatures and a deconfined quark-gluon plasma (QGP) at very high temperatures. Hot QCD, as observed at RHIC
and LHC, is a highly collective and strongly interacting medium with a very small mean free path of the effective constituents \cite{Heinz:2013th}. It has been established on the lattice that this matter in equilibrium is not only chirally symmetric but also approximately chiral spin symmetric~\cite{Rohrhofer:2017grg,Rohrhofer:2019qwq,Rohrhofer:2019qal,Chiu:2023hnm}.
The chiral spin symmetry $SU(2)_{CS}$   \cite{Glozman:2014mka,Glozman:2015qva} 
is a symmetry of the color charge and confining electric part of the QCD Lagrangian (for a review see Ref.~\cite{Glozman:2022zpy}).  For an arbitrary number of quark flavors $N_f$, it can be extended to $SU(2N_F)$, that includes the chiral group
$SU(N_F)_R \times SU(N_F)_L \times U(1)_A$ as a subgroup. Observation of this
symmetry above $\Tch$ suggests that QCD is still in the confining
regime.  The regime of QCD between the chiral restoration crossover and a very smooth crossover to QGP was tagged as stringy fluid. There exist further lattice evidences, not related to symmetries, supporting
the existence of this intermediate regime of QCD above $\Tch$ \cite{Glozman:2022lda,Lowdon:2022xcl,Bala:2023iqu,hot,Allton}.
The energy density, pressure and entropy density within three different QCD regimes scale as $N_c^0$ in the hadron gas, as $N_c^1$ in the stringy fluid
and as $N_c^2$ in the  QGP \cite{Cohen:2023hbq,Cohen:2024ffx,Fujimoto:2025sxx}.

In this paper we clarify the origin of the $N_c^1$ scaling
within the stringy fluid.
We employ a solvable manifestly confining and chirally symmetric large $N_c$ quark model in 3+1 dimensions  \cite{Amer:1983qa,LeYaouanc:1983huv,LeYaouanc:1984ntu,Adler:1984ri,Kocic:1985uq,Bicudo:1989sh,Bicudo:1989si,Llanes-Estrada:1999nat,Bicudo:2002eu,Nefediev:2004by,Alkofer:2005ug,Wagenbrunn:2007ie,Glozman:2007tv,Glozman:2009sa}. The Hamiltonian of the model contains only
the quark-kinetic term and the electric confining linear potential between the color
charge densities  of quarks. This model shares the chiral spin symmetry of the confining interaction with QCD \cite{Glozman:2024xll}.
The model is reminiscent of the celebrated 't~Hooft model for QCD in 1+1 dimensions in the large-$N_c$ limit \cite{tHooft:1974pnl}. 

Earlier, together with Nefediev and Wagenbrunn, we derived and solved the finite-temperature mass-gap equation in this model and observed
the chiral symmetry restoration phase transition at a temperature $\Tch \sim 90 $
MeV \cite{Glozman:2024xll,GNW}.
\footnote{A similar result was obtained in Ref.~\cite{Quandt:2018bbu} in a different framework based on compactification of one spatial direction in Euclidean space.}. We have also clarified the physical mechanism of the chiral symmetry restoration: the thermal excitations of quarks and antiquarks
lead to Pauli blocking of the levels necessary for the formation of the quark condensate. Since the confining potential is assumed to be temperature
independent the model should adequately address qualitative features of the
stringy fluid phase between $T_{ch}$ and  $T_d >> T_{ch}$. We have studied properties of the bound quark-antiquark states
at small temperatures as well as at $T > T_{ch}$ and observed that
\begin{itemize}
\item[(i)] above $\Tch$ the spectrum of the quark-antiquark excitations exhibits chiral symmetry and approximate $SU(4) \times SU(4)$ symmetry of confinement
for $J > 0$ ;
\item[(ii)] the light-light quark-antiquark color-singlet low-spin excitations acquire a much larger size at $T>\Tch$ than at $T<\Tch$;  in the strict chiral limit the low-spin meson-like
systems become infinitely large even though they are still in the confining regime.
\end{itemize}
The latter property suggests that the stringy fluid matter is a dense system of long and strongly overlapping "strings". This insight may facilitate an
explanation of the experimental observations made for QCD above $\Tch$: a collectivity and a very small mean free path of the effective constituents.
This paper  relies on the  results obtained in refs. \cite{Cohen:2023hbq,Cohen:2024ffx,Glozman:2024xll,GNW} and
presents the reason for the $N_c^1$ scaling of the energy density in the stringy fluid.

\section{Confining and chirally symmetric model in 3+1 dimensions}

The model Hamiltonian is an approximation to the QCD Hamiltonian in the Coulomb
gauge and its gluonic part retains only  the instantaneous confining ``Coulombic'' term 
\be
\begin{split}
H=&\int d^3x\;\psi^\dagger(\vex,t)\left(-i\vec{\alpha}\cdot
{\bm\nabla}+\beta \mq\right)\psi(\vex,t)\\
& + \frac{1}{2} \int d^3x\; d^3y\;\rho^a(\vex)K_{ab}(|\vex-\vey|)\rho^b(\vey),
\label{GNJL}
\end{split}
\ee
which includes the interaction of two quark color charge densities,
$\rho^a=\psi^\dag\frac{\lambda^a}{2}\psi$, taken at the spatial points $\vex$ and $\vey$, via an instantaneous confining kernel,
\be
K_{ab}(|\vex-\vey|)=\delta_{ab}V_0(|\vex-\vey|).
\label{Kab}
\ee
The quark kinetic part is chirally symmetric while the confining ``Coulombic'' part is invariant under larger symmetry groups: $SU(2)_{CS}$, $SU(2N_F)$, and $SU(2N_F) \times SU(2N_F)$ \cite{Glozman:2022zpy}.
A standard approach to solving the model implies a rainbow approximation for the dressed quark Green's function and a ladder approximation for the quark-antiquark Bethe--Salpeter equation. Such approximation is well justified in the large-$N_c$ limit.  
A linearly rising confinement potential is
\be
V_{\rm conf}(r)= C_F V_0(r)=\sigma r,
\label{Vlin}
\ee
with $C_F$ being the color Casimir factor and $\sigma$ represents the fundamental "Coulomb string tension". The $N_c$-scaling of $V_0(r)$ must be such that
$\sigma$ reaches a finite limit at $N_c \rightarrow \infty$. This is necessary
to achieve a consistency with the 't Hooft $N_c$ conting rules \cite{tHooft:1973alw}
and in particular with the $N_c^0$ scaling of meson masses \cite{tHooft:1973alw,witten}.

 In Refs.~\cite{Szczepaniak:2001rg,Feuchter:2004mk}, the linear confinement was obtained in variational calculations in quenched Coulomb-gauge QCD. Notice that the "Coulomb string tension" is larger than the string tension associated with the area law of the Wilson loop \cite{Z}.
Another attractive feature of the considered model with the linear potential \eqref{Vlin} is its direct analogy with the 't~Hooft model for QCD in 1+1 dimensions in the large-$N_c$ limit \cite{tHooft:1974pnl} that was extensively studied in the axial (Coulomb) gauge, for example, in Refs.~\cite{Bars:1977ud,Li:1986gf,Kalashnikova:2001df,Glozman:2012ev}. In the two-dimensional model, the appearance of a linearly rising potential between quarks is a natural consequence of the form of the two-dimensional gluon propagator in the Coulomb gauge. 

\begin{figure*}[t!]
\centering
\includegraphics[width=0.9\textwidth]{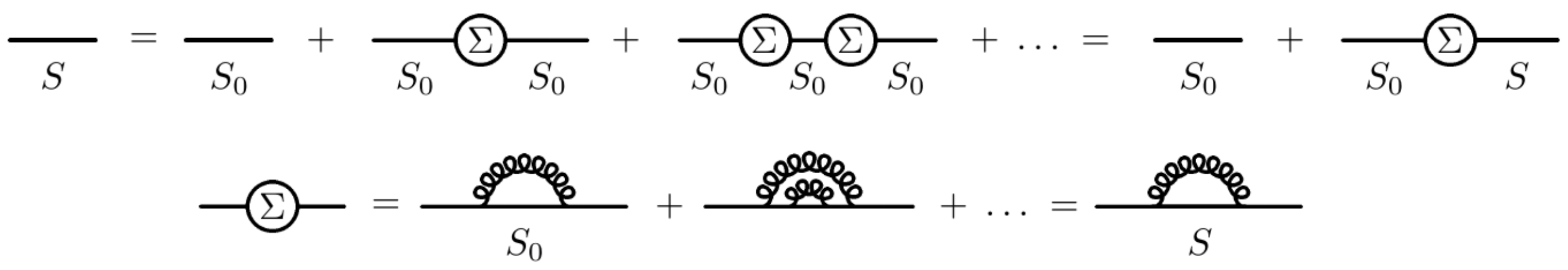}
\caption{Schematic representation of the gap equation derived in the rainbow approximation. The thin and thick solid lines are for the bare and dressed quark propagator, respectively, and the curly line is for the confining gluon propagator.}
\label{fig:dyson}
\end{figure*}

\begin{figure}[t!]
\centering
\includegraphics[width=0.75\columnwidth]{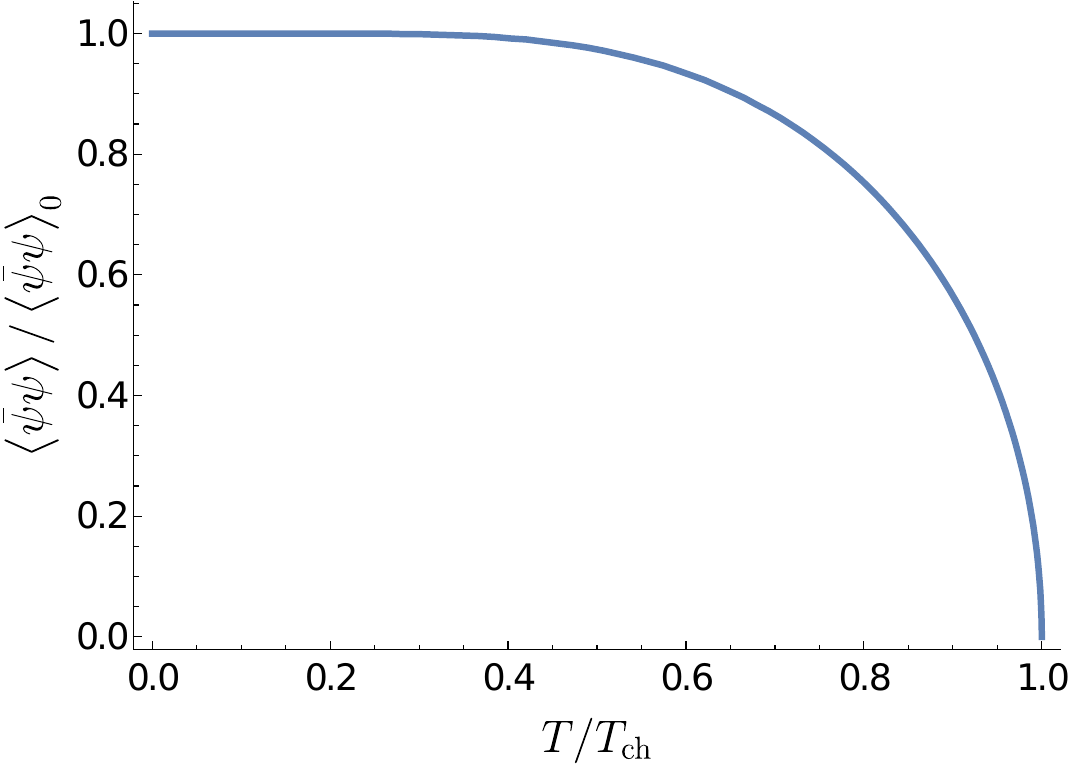}
\caption{Temperature dependence of the chiral condensate  normalized to its maximum value reached at $T=0$.  Adapted from Ref.~\cite{Glozman:2024xll}.}
\label{fig:cond}
\end{figure}

With a confining interquark interaction, the trivial chirally-symmetric vacuum is unstable. This instability can be studied with the help of the Bogoliubov-Valatin transformation of the quark field, which is equivalent to the solution of the Schwinger-Dyson
gap equation in the rainbow approximation, see
Fig. \ref{fig:dyson}.
The gap equation with the linear confining potential  leads to the chiral symmetry breaking solution in the vacuum
(it was first solved in Ref. \cite{Adler:1984ri}), 
\be
 \braket{\bar{\psi}\psi}_{T=0}\approx-(0.23\sqrt{\sigma})^3.
\label{cond1}
\ee
 The finite temperature gap equation was obtained and solved in refs.
\cite{Glozman:2024xll,GNW}. At the critical temperature $T_{ch}$ the chiral
symmetry gets restored and the chiral condensate vanishes,
\be
\Tch\approx 0.084\sqrt{\sigma},
\label{Tch}
\ee
see Fig. \ref{fig:cond}.
For the phenomenological value of the chiral condensate $\braket{\bar{\psi}\psi}_0=-(250~\mbox{MeV})^3$, it predicts
\be
\Tch\approx 90~\mbox{MeV},
\ee
which is a decent estimate for the lattice chiral phase transition temperature
in the chiral limit , $\Tch \simeq 130$ MeV \cite{HotQCD:2019xnw}, given a very simple form of the interquark interaction employed in the calculation.
The physical reason of the chiral restoration in the confining regime is Pauli
blocking of the quark and antiquark levels, required for the existence of a nonvanishing
quark condensate, by the thermal excitation of quarks and antiquarks.

\begin{figure}[t!]
\centering
\includegraphics[width=0.8\columnwidth]{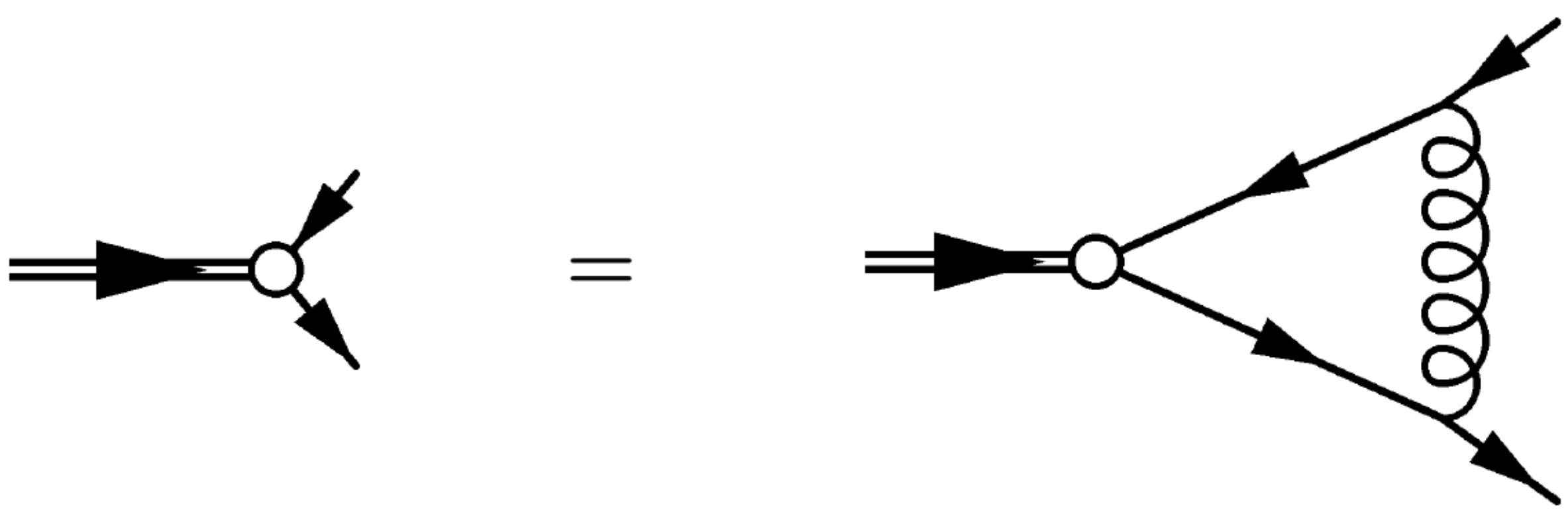}
\caption{The Bethe--Salpeter equation  written in the ladder approximation for the interquark interaction. The single, double, and curly line correspond to the quark (antiquark), meson, and the interquark interaction, respectively.}
\label{fig:bseq}
\end{figure}
Given the single quark Green function, obtained from the gap
equation, one can address the quark-antiquark bound states via
solution of the Bethe-Salpeter equation 
in the ladder approximation,
see Fig.~\ref{fig:bseq}.  The Bethe-Salpeter equation in the quark-antiquark rest frame
for the matrix amplitude $\chi(\vep;M)$, with $\vep$ for the momentum of the quark ($-\vep$ for the antiquark) and $M$ for the mass of the meson,  reads
\cite{Wagenbrunn:2007ie}
\be
\begin{split}
\chi({\vec p};M)=i\int&\frac{d^4q}{(2\pi)^4}V(\vep-\veq)\gamma_0 S(q_0+M/2,\veq)\\
&\times\chi(\veq;M)S(q_0-M/2,\veq)\gamma_0,
\label{BSeq}
\end{split}
\ee
where the propagator $S(p_0,\vep)$ is the single
quark Green function obtained from the gap equation. The regularized
in the infrared confining gluon propagator $V(\vep)=-V_{conf}(\vep)$ is given as
\be
V(p)=\frac{8\pi\sigma}{(p^2+\muIR^2)^2}.
\label{FV}
\ee
Then, in coordinate space, we have
\be
V(r)=\int \frac{d^3 p}{(2\pi)^3} V(p) e^{i\vec p \vec r}=\frac{\sigma}{\muIR}e^{-\muIR r}.
\label{div}
\ee
It is easy to see that, in the limit of $\muIR\to 0$,
\be
V_{\rm conf}(r)=-V(r)\mathop{=}_{\muIR\to 0}-\frac{\sigma}{\muIR}+\sigma r+\ldots,
\label{Vreglim}
\ee
where the ellipsis stands for the terms that vanish in the infrared limit
$\mu_{IR}=0$. Notice that within this manifestly confining model
a color triplet single quark does not exist on-mass-shell because the single
quark Green function is infrared-divergent and there are no single
quark poles in the complex energy plane. However, 
the infrared-divergent constant $-\frac{\sigma}{\muIR}$ cancels in the
Bethe-Salpeter equation for a color-singlet quark-antiquark system because the divergence of the quark Green functions cancels with the
divergence of the Bethe-Salpeter kernel \cite{Wagenbrunn:2007ie}.  Such cancellation happens in any color-singlet
system \cite{Glozman:2008fk}.

The effect of a finite temperature is accommodated in the quark
propagator, obtained from the finite temperature gap equation, as well
as via a replacement
$V(\vec p - \vec q) \to (1-n_q-\bar{n}_q)V(\vec p - \vec q)$ in the Bethe-Salpeter equation, where the Fermi-Dirac distribution functions for quarks and antiquarks, $n_q$ and $\bar{n}_q$,
contain the infrared-finite dynamical quark mass $M_q$ obtained from the solution of the thermal mass-gap equation at the given finite temperature $T$ \cite{GNW}:
\begin{align}
n_q=\left(1 + e^{(\sqrt{q^2 + M_q^2}-\mu)/T}\right)^{-1},\nonumber\\[-2mm]
\label{nnnew}\\[-2mm]
\bar{n}_q=\left(1 + e^{(\sqrt{q^2 + M_q^2}+\mu)/T}\right)^{-1}.\nonumber
\end{align}
It implies that, at all temperatures, the effective interaction in the quark--antiquark meson is exactly the same as in the mass-gap equation.

\section{Properties of the quark-antiquark excitations at different temperatures.}

The Bethe-Salpeter equation at temperatures below and above $T_{ch}$
was solved in Ref. \cite{GNW} and the spectra
as well as the properties of the bound quark-antiquark systems have been studied at
different temperatures.   
 Here we overview a property of the Bethe-Salpeter
"wave functions" that is relevant to the subject of the present note.

The solution of the Bethe-Salpeter equation contains the propagating forward
and backward in time quark-antiquark
"wave functions" $\psi_+(p)$ and $\psi_-(p)$ that are normalized according to
\be
\int\frac{p^2dp}{2\pi^2}\Bigl[\psi_+^2(p)-\psi_-^2(p)\Bigr]=2M.
\label{norm2M}
\ee
The relativistic normalization in Eq.~\eqref{norm2M} is consistent with the condition that the charge of the state with $I=I_3=1$ equals to unity.

\begin{figure*}[t!]
\centering
\includegraphics[width=0.5\textwidth]{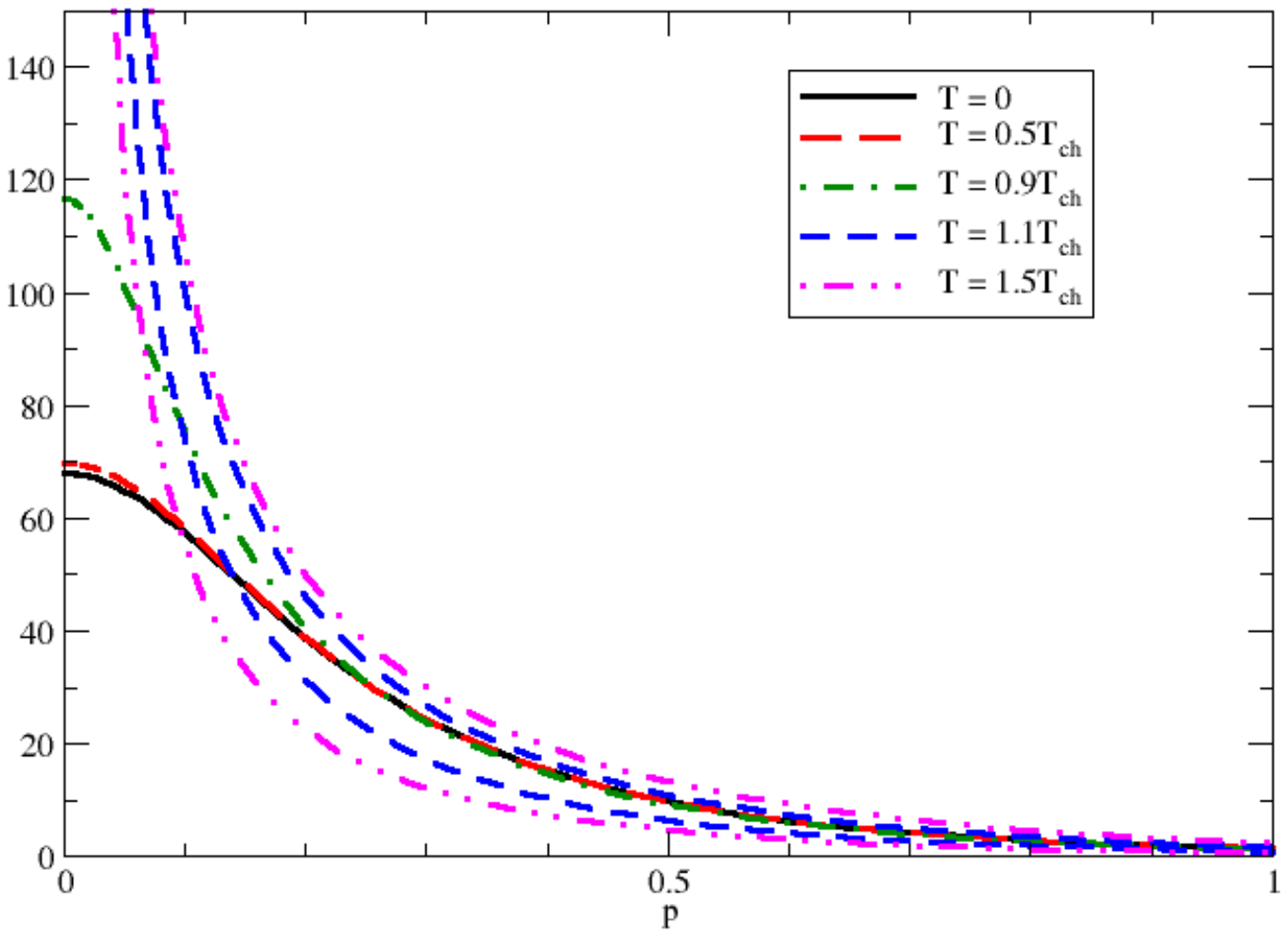}
\hspace*{-0.02\textwidth}\includegraphics[width=0.5\textwidth]{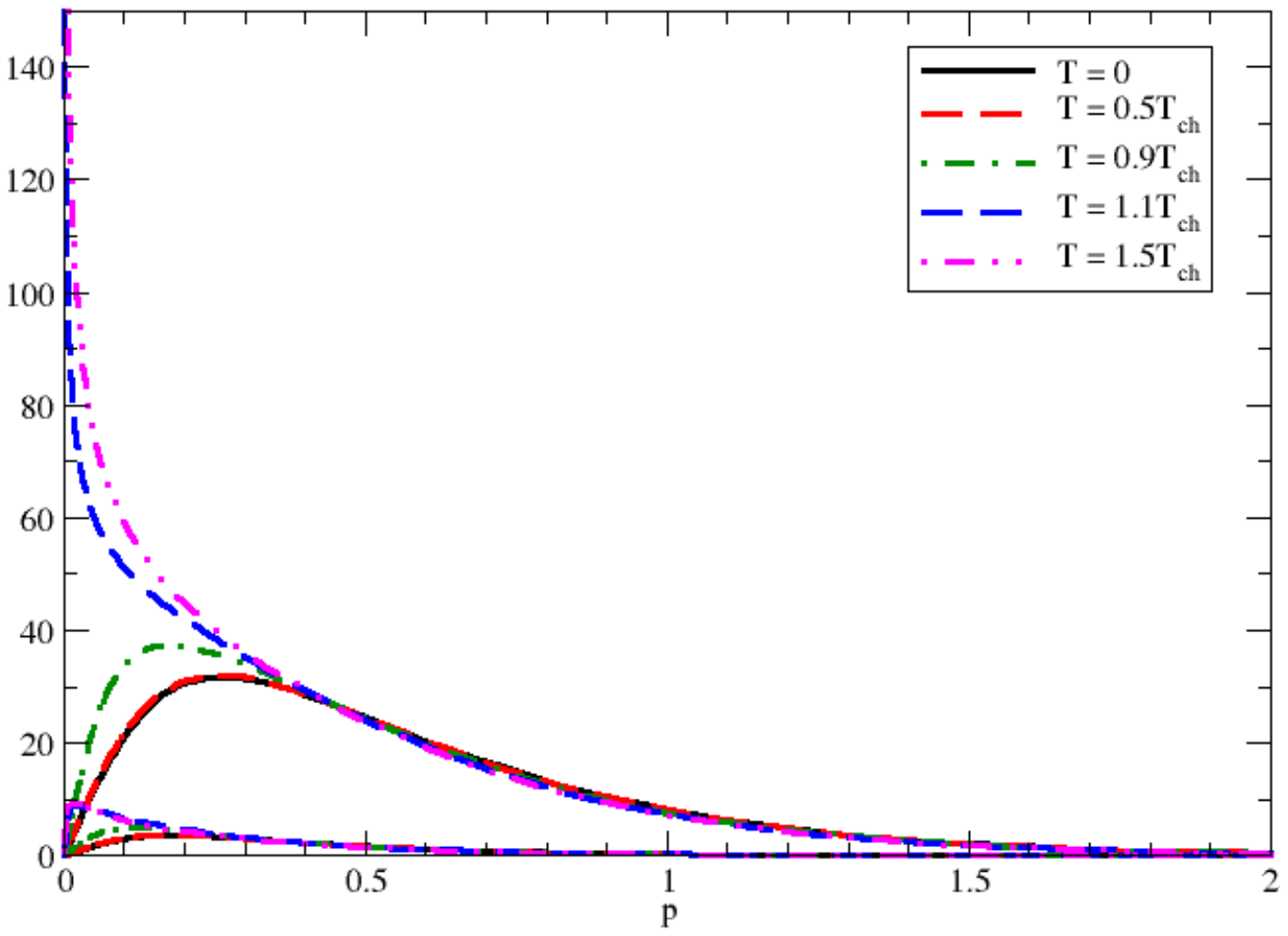}
\caption{The ground state ($n=0$) wave functions $\psi_\pm(p)$ for $J^{PC}=0^{-+}$ ( left plot), $J^{PC}=1^{+-}$ (right plot)  at different temperatures. For $J^{PC}=0^{-+}$ and $T<\Tch$, the corresponding pseudoscalar meson is a massless Goldstone boson with $\psi_+(p)=\psi_-(p)$. For $J^{PC}=0^{-+}$ and $T>\Tch$ as well as for all other sets of quantum numbers and all temperatures, $\psi_+(p)>\psi_-(p)$, so for each temperature there are two curves in the corresponding plots. All dimensional quantities are given in the
appropriate units of $\sqrt \sigma$. Adapted from Ref. \cite{GNW}.}
\label{fig:wfs}
\end{figure*}

In Fig. \ref{fig:wfs} we demonstrate the wave functions  $\psi_+(p)$ and $\psi_-(p)$ for the ground color-singlet  quark-antiquark systems with quantum numbers $J^{PC}=0^{-+}$ and $J^{PC}=1^{+-}$
obtained from the solution of the Bethe-Salpeter equation at different temperatures in the chiral limit
$m_q \rightarrow 0$.
At $T=0$ and all temperatures $T < T_{ch}$ the meson wave functions with low
spin $J$ are
localized in a small space volume. This is a result of a confining interaction
between quarks that acquire a dynamical mass $M_q$ due to the spontaneous breaking
of chiral symmetry. This localization is reflected in a small root-mean-square
radius of mesons \cite{GNW}. At the chiral restoration temperature
$T_{ch}$ the thermal excitations of quarks and antiquarks block the levels
required for the existence of a non-vanishing quark condensate and chiral
symmetry gets restored. This Pauli blocking implies an effective modification
of the confining potential (\ref{FV}), because the Fermi-Dirac distributions
of quarks and antiquarks above $T_{ch}$ give
\be
(1-n_q-\bar{n}_q)_{|\mu=0,M_q=0,T>0}\mathop{\propto}_{q\to 0}q.
\label{PB}
\ee
Consequently the effective potential that determines the meson wave function in the Bethe-Salpeter equation, $(1-n_q-\bar{n}_q)V(\vec p - \vec q)$, loses
its confining properties and the color-singlet quark-antiquark system
gets delocalized. This delocalization can be seen from the fact that
the wave functions $\psi_+(p),\psi_-(p)$ are divergent at $p=0$ and the
system acquires infinite root-mean-square radius in the chiral limit.
At realistic quark masses the size of the low-spin mesons increases
by a factor $\sim 5$ as compared to their size below $T_{ch}$ \cite{GNW}. Notice that
the delocalized color-singlet quark-antiquark system still knows about confinement: the color-triplet free quarks do not exist and the spectrum
of the color-singlet  quark-antiquark systems is discrete \cite{GNW}.

The huge swelling of low-spin "mesons" above $T_{ch}$ has significant
phenomenological implications. The hot QCD matter above $T_{ch}$ but below
$T_d$ is a dense medium of overlapping huge color-singlet quark-antiquark
systems ("strings"). Consequently the matter is highly collective and the mean free path of the effective color-singlet constituents approaches zero.

\section{The $N_c^1$ scaling in the stringy fluid.}

Now we are in a position to address the $N_c^1$ scaling above $T_{ch}$
but below $T_d$, which is the aim of the present note. In  the hadron gas phase below
$T_{ch}$ the thermodynamic quantities scale as $N_c^0$.
This is because the quark-antiquark meson mass in the vacuum scales as $N_c^0$
\cite{witten}. The physical reason for this scaling is the strong color-exchange interaction between the quark and antiquark that leads to confinement and
to a compact size of mesons. This feature is well seen within the 't Hooft model
that provides the same scaling of meson masses.

Given that mesons in the hadron gas do not interact at large $N_c$,
the Bose-Einstein distribution for the meson 
number density of species $k$ is valid:
\begin{equation}
  n_k(T)  = (2S_k+1)(2I_k+1) \int \frac{{\rm d}^3 p}{(2 \pi)^3} \, \frac{1}{e^{\sqrt{p^2 + m_k^2}/T} - 1}  \; .
\label{num}
\end{equation}
Since the meson masses are of the order $N_c^0$, one immediately obtains
that the meson number density in the hadron gas scales as $N_c^0$. The energy
density is given by the product of the number density and meson energy
and consequently also scales as $N_c^0$.

 Above $T_{ch}$ the situation changes qualitatively.
What is the scaling of the meson-like excitation energies above $T_{ch}$?
\footnote{ Note that in a dense medium at high temperatures the 
Lorentz invariance is lost and the rest-frame  excitation energy in the
color-singlet quark-antiquark system cannot be identified with the meson mass.} 
The Bethe-Salpeter equation (\ref{BSeq}) does not contain
the $N_c$ dependent factors \cite{GNW}\footnote{The author is thankful to A. Nefediev for stressing this point.}. 
The same is also true  in the 't Hooft model where the $N_c$-factor is
absorbed into the 't Hooft coupling that is kept constant and finite.
So the Bethe-Salpeter equation in the axial (Coulomb) gauge as well as
the 't Hooft bound state equation in the light cone gauge are $N_c$-independent,
see, e.g., ref. \cite{Glozman:2012ev}.
Consequently the spectrum of  the excitation energies in the color-singlet
quark-antiquark systems with different quantum numbers $k$, obtained in ref. \cite{GNW},
in vacuum
as well as at finite temperatures, including $T > T_{ch}$, does not depend on 
$N_c$: the excitations energies $E_k$ scale as $N_c^0$ not only in the vacuum and in the hadron gas,
but also above $T_{ch}$.  
This scaling of the excitation energies above $T_{ch}$ is the only output
of the discussed model that is used to reach the conclusion given below.
The conclusion is general and does not depend on model details.

The medium above $T_{ch}$ is strongly interacting and the ideal gas Bose-Einstein distributions, that are  used for the hadron gas below $T_{ch}$, are not applicable. At vanishing chemical potential all expectation values 
of the bilinear operators $\bar q \Gamma_k q$ (where $\Gamma_k$ includes $\gamma$- and isospin-matrices) with not vacuum quantum numbers automatically vanish,
$<\bar q \Gamma_k q>=0$.
Consequently they can not be used as a substitute of the meson number density 
(\ref{num})
in a dense medium. However, the fluctuations (variance) of these quantities,
do not vanish and can be used to evaluate the energy density above $T_{ch}$.
So we evaluate the energy density in a dense medium as a product  of the  
 square root of fluctuations   of such color-singlet quark-antiquark pairs,
 the standard deviation
 \begin{equation}
 \sigma_k =\sqrt{\int d^3 x_1 d^3 x_2 <\bar q(x_1) \Gamma_k q(x_1) \bar q(x_2) \Gamma_k q(x_2)>}
 \label{fl}
\end{equation}
 (this linear measure of fluctuations should substitute (\ref{num}) in a dense medium)
 and the energy  $E_k$ of each pair:
\begin{equation} 
\sum_k \sigma_k E_k  
\label{de}.
\end{equation}
We do not yet know the $N_c$ scaling of the fluctuations above
$T_{ch}$. Consider, as example, the conserved charges
which are defined as expectation values of  specific bilinears
\begin{equation}    
N \equiv \int d^3 x ~n(x) \; \;\;  {\rm with}
\; \; \;n(x) = \bar q(x) \gamma^0 q(x), \;\; \; q=u,d
\label{def}.
\end{equation}
These expectation values  do vanish in the system with vanishing
chemical potential. However, their fluctuations do not.
 The important issue is that the linear measure  of the fluctuations of conserved charges
 scale 
as $N_c^1$ in the confining but chirally symmetric
phase above $T_{ch}$. This is because the
bilinear $n(x)$ in (\ref{def}) scales as $N_c^1$ \cite{Cohen:2024ffx}. Note that the fluctuations of conserved charges in the hadron gas, obtained from (\ref{num}), scale
as $N_c^0$.
It is rather obvious that the fluctuations of other quark bilinears in a dense medium scale as
of conserved charges. Hence,
one obtains that the energy density (\ref{de}) in the stringy fluid phase scales
as $N_c^1$.
This picture must  obviously be
checked in explicit lattice calculations.

 We conclude that 
 the energy density of the hot matter with confinement above
$T_{ch}$ but below $T_d$ scales as $N_c^1$, i.e.,  as if it were a gas of quasi-free quarks.
This scaling prescribes the same scaling of pressure and entropy density.

\section{Conclusions}
In this work we have discussed the origin of the $N_c^1$ scaling
of the thermodynamic quantities in the stringy fluid phase at infinite $N_c$
in the chiral limit.
 We have employed a manifestly
confining and chirally symmetric large $N_c$ field theoretical model in
3+1 dimensions that is similar to 't Hooft model in 1+1 dimensions. The
linear confining potential between the color charge densities of quarks
induces the spontaneous breaking of  chiral symmetry in the vacuum and the localization of quarks and antiquarks in a very small space volume of mesons. At
the chiral restoration temperature the chiral symmetry gets restored because
of  Pauli blocking of the quark levels necessary for the formation of the
quark condensate, by the thermal excitations of quarks and antiquarks. The
same Pauli blocking leads to the infinite delocalization of the low-spin mesons that
 represent a superposition of the
forward and backward in time propagating color-singlet quark-antiquark
pairs.  The
stringy fluid matter represents a dense medium of huge strongly overlapping
color-singlet quark-antiquark low-spin systems  with the $N_c^1$ scaling of energy density, pressure and entropy density. It is a highly collective medium
 with a very small
mean free path of the  color-singlet constituents.

\section*{Acknowledgments}
The author is thankful to T. Cohen and A. Nefediev for constructive
discussions.
This work was supported by the Austrian Science Fund (FWF) through the grant
PAT3259224.

\end{document}